\documentclass{article}

\usepackage{graphicx}
\def\tbzresult{1.50\pm0.05\pm0.07}
\def\tbmresult{1.70\pm0.06{}^{+0.11}_{-0.10}}
\def\rbmresult{1.14\pm0.06^{+0.06}_{-0.05}}
\def\yresult{0.03_{-0.18}^{+0.15}{}_{-0.08}^{+0.05}}
\def\yllimit{-0.36}
\def\yulimit{0.35}
\def\tbzdstlnu{1.50\pm0.06^{+0.06}_{-0.04}}
\def\tbmdstlnu{1.54\pm0.10^{+0.14}_{-0.07}}
\def\tbzdstpm{1.55^{+0.18}_{-0.17}{}^{+0.10}_{-0.07}}
\def\tbzdppm{1.41^{+0.13}_{-0.12}\pm0.07}
\def\tbmdzpm{1.73\pm0.10\pm0.09}
\def\tbmpsikm{1.87^{+0.13}_{-0.12}{}^{+0.07}_{-0.14}}
\def\tbzpsiks{1.54^{+0.28}_{-0.24}{}^{+0.11}_{-0.19}}
\def\tbzpsikst{1.56^{+0.22}_{-0.19}{}^{+0.09}_{-0.15}}
\def\tdzresult{414.8\pm3.8\pm3.4}
\def\tdpresult{1040^{+23}_{-22}\pm18}
\def\tdsresult{479^{+17}_{-16}{}^{+6}_{-8}}
\def\rdpresult{2.51\pm 0.06\pm0.04}
\def\rdsresult{1.15\pm 0.04^{+0.01}_{-0.02}}
\def\ydresult{1.0^{+3.8}_{-3.5}{}^{+1.1}_{-2.1}}
\def\ydllimit{-7.0}
\def\ydulimit{8.7}

\def\Tbzresult{$\taubz=(\tbzresult)$~ps}
\def\Tbmresult{$\taubm=(\tbmresult)$~ps}
\def\Rbmresult{$\rbm=\rbmresult$}
\def\Yresult{$\ycp=\yresult$}
\def\Ylimit{$\yllimit<\ycp<\yulimit$}
\def\tdz{\tau(\dz)}
\def\tdp{\tau(\dplus)}
\def\tds{\tau(\ds)}
\def\rdp{\tdp/\tdz}
\def\rds{\tds/\tdz}
\def\Tdzresult{$\tdz=(\tdzresult)$~fs}
\def\Tdpresult{$\tdp=(\tdpresult)$~fs}
\def\Tdsresult{$\tds=(\tdsresult)$~fs}
\def\Rdpresult{$\rdp=\rdpresult$}
\def\Rdsresult{$\rds=\rdsresult$}
\def\Ydresult{$\ycp=(\ydresult)$\%}
\def\Ydlimit{$\ydllimit\%<\ycp<\ydulimit$\%}

\newcommand{\zpc}[3]{Z. Phys. C {\bf #1} (#3) #2}

\def\bzb{{\overline{B}^0}}

\def\bm{{B^-}}
\def\taubz{\tau(\bzb)}
\def\taubm{\tau(\bm)}
\def\rbm{\taubm/\taubz}
\def\Bzb{$\bzb$}
\def\Bm{$\bm$}

\def\ycp{y_{CP}}
\def\piz{\pi^0}
\def\pip{\pi^+}
\def\pim{\pi^-}

\def\kp{K^+}
\def\km{K^-}
\def\ks{K_S}
\def\kb{\overline{K}}

\def\dstar{D^{*}}
\def\dstarz{{D^{*0}}}
\def\dstarp{D^{*+}}

\def\Dstar{$\dstar$}

\def\nub{\overline{\nu}}
\def\jpsi{{J/\psi}}
\def\dz{D^0}
\def\Dz{$\dz$}
\def\dplus{D^+}
\def\Dp{$\dplus$}
\def\dzbar{\overline{D}^0}

\def\ds{D_s^+}
\def\Ds{$\ds$}

\def\bdstlnu{\overline{B}\to\dstar\ell^-\nub}
\def\bdstxlnu{\overline{B}\to\dstar X\ell^-\nub}
\def\bzdstlnu{\bzb\to\dstarp\ell^-\nub}

\def\bmdstlnu{\bm\to\dstarz\ell^-\nub}

\def\bpsik{\overline{B}\to\jpsi\kb}

\def\bzdstpm{\bzb\to\dstarp\pim}
\def\bzdppm{\bzb\to\dplus\pim}
\def\bmdzpm{\bm\to\dz\pim}

\def\bzbpsiks{\bzb\to\jpsi\ks}

\def\bzbpsikst{\bzb\to\jpsi\kstarzb}
\def\bmpsikm{\bm\to\jpsi\km}

\def\kmkp{\km\kp}
\def\kpi{\km\pip}

\def\kpipi{\km\pip\pip}
\def\kpipiz{\km\pip\piz}
\def\kpipipi{\km\pip\pip\pim}
\def\kstarzb{\overline{K}^{*0}}

\def\dzkpi{\dz\to\kpi}
\def\dzkk{\dz\to\kmkp}
\def\dpphipi{\dplus\to\phi\pip}
\def\dpkpipi{\dplus\to\kpipi}
\def\dsphipi{\ds\to\phi\pip}
\def\dskstk{\ds\to\kstarzb\kp}

\def\chisq{\chi^2}

\def\Chisq{$\chisq$}

\def\dM{\Delta M}

\def\amix{A_{mix}}
\def\re{{\cal R}e}

\def\dG{\Delta\Gamma}

\def\dE{\Delta E}
\def\DE{$\dE$}
\def\mb{M_b}
\def\Mb{$\mb$}
\def\delz{\Delta z}
\def\dt{\Delta t}

\def\dtb{\dt_i}
\def\dtp{\dt'}

\def\dtrec{\dt_{rec}}
\def\dtgen{\dt_{gen}}

\def\sigdt{\sigma_i}

\def\sigmisdt{\sigma_{t}^i}
\def\sigdz{\sigma_{\delz}}

\def\sigzrec{\sigma_{z}^{rec}}
\def\sigzasc{\sigma_{z}^{asc}}

\def\sigk{\sigma_{K}}

\def\smis{S_{t}}
\def\scharm{S_{charm}}
\def\sdet{S_{det}}

\def\smisbg{S_{t}^{BG}}
\def\sbg{S_{BG}}
\def\mumis{\mu_{t}}

\def\mumisbg{\mu_{t}^{BG}}

\def\mubg{\mu_{BG}}
\def\fmis{f_{t}}
\def\fmisbg{f_{t}^{BG}}
\def\flmbg{f_{\lambda BG}}

\def\fbg{f_{BG}}

\def\lmbg{\lambda_{BG}}

\def\fm{f_-}
\def\fz{f_0}

\def\tbm{\tau_-}
\def\tbz{\tau_0}
\def\tsig{\tau_{sig}}

\def\etal{{\it et al.}}

\def\micron{$\mu$m}

\begin{document}

\vspace*{4cm}
\begin{center}
{\LARGE\bf Measurements of Heavy Meson Lifetimes with Belle}

\bigskip
\bigskip
\bigskip

{\large H. Tajima\\
(Belle Collaboration)\\
\smallskip
Department of Physics, University of Tokyo, 7-3-1 Hongo, Tokyo 113-0033 Japan\\
 E-mail: tajima@phys.s.u-tokyo.ac.jp}\\
 
\bigskip
\bigskip
\bigskip
\bigskip

{\large\bf Abstract}
\end{center}


\smallskip
Charmed and beauty meson lifetimes have been measured using 
$2.75$ fb$^{-1}$ ($D$ mesons) and $5.1$ fb$^{-1}$ ($B$ mesons) of 
data collected with the Belle detector at KEKB.
The results are \Tbzresult, \Tbmresult, 
\Tdzresult, \Tdpresult\ and \Tdsresult, where the first error is 
statistical and the second error is systematic. 
The lifetime ratios are measured to be \Rbmresult, \Rdpresult\ and \Rdsresult. 
The mixing parameter $\ycp$ is also measured to be \Yresult\ 
for \Bzb\ and \Ydresult\ for \Dz, corresponding to 
95\% confidence intervals, \Ylimit\ and \Ydlimit, respectively.
All results are preliminary.

\bigskip
\bigskip

\begin{center}
Contributed to the Proceedings of the 30th International Conference on High Energy Physics, \\
July 27 -- August 2, 2000, Osaka, Japan.
\end{center}
\thispagestyle{empty}

\twocolumn
Measurements of individual heavy meson lifetimes provide 
useful information for the theoretical understanding of 
heavy meson decay mechanisms. 
In particular, experimental results\cite{HF99} yield 
$\tau(\ds)/\tau(\dz)=1.191\pm0.024$, which is 
inconsistent with the theoretically expected range\cite{life-ratio-th}
of 1.00--1.07.
Moreover, measurements of the differences of lifetimes for
neutral mesons decaying into CP-mixed states and CP-eigenstates
can be used to study the $y\equiv\dG/2\Gamma$  
and $x\equiv \Delta M/\Gamma$ particle-antiparticle mixing parameters.

The parameter $\ycp$, defined as
\begin{eqnarray*}
\ycp&\equiv&\frac{\Gamma(\mathrm{CP\ even})-\Gamma(\mathrm{CP\ odd})}{\Gamma(\mathrm{CP\ even})+\Gamma(\mathrm{CP\ odd})}
\end{eqnarray*}
is related to $y$ and $x$ by the expression
\begin{eqnarray*}
\ycp&=&\frac{\tau(\dzkpi)}{\tau(\dzkk)}-1\\
&\approx&y\cos\phi-\frac{\amix}{2} x\sin\phi,\\
\ycp&=&1-\frac{\tau(\bzdstlnu,\bzb\to D\pi)}{\tau(\bzbpsiks)}\\
&\approx&y\cos2\phi_1,
\end{eqnarray*}
where $\phi$($\phi_1$) is a CP-violating 
weak phase due to the interference of decays 
with and without mixing, and $\amix$ is a state-mixing CP-violating 
parameter ($\amix\approx4\re(\epsilon)$). 
The FOCUS experiment reports $\ycp=(3.42\pm1.39\pm0.74)$\%\cite{FOCUS}, 
while CLEO gives $y'\cos\phi=(-2.5^{+1.4}_{-1.6})$\%\cite{CLEOb}, 
$x'=(0.0\pm1.5\pm0.2)$\% and 
$\amix=0.23^{+0.63}_{-0.80}$ using $\dz\to\kp\pim$, 
where $y'=y\cos\delta - x\sin\delta$ and $x'=x\cos\delta + y\sin\delta$; 
$\delta$ is a strong phase between $\dz\to\kp\pim$ 
and $\dzbar\to\kp\pim$ decays.
These results may be an indication of a large $SU(3)$-breaking 
effect in $\dz\to K^\pm\pi^\mp$ decays\cite{mix-th}.

This report mainly describes the $\bdstlnu$ analysis.  
The $D$ lifetime analyses are described in Ref. \ref{ref:belle-dlife}.


Candidate $\bdstlnu$ decays are selected by applying 
kinematic constraints on events with a lepton and a 
$\dstar\to \dz\pi$ decay chain, where $\dz\to\kpi$, $\kpipiz$ and $\kpipipi$
decays are used.  First, the
\Dz\ decay vertex is determined and then
the decay vertex of the $\bdstlnu$ candidate is calculated using
the lepton and the inferred \Dz\ track.  
The vertex point of the accompanying $B$ meson 
is determined from the remaining tracks, after the rejection 
of $K_S$ daughters and badly measured tracks.
When the reduced \Chisq\ of the vertex fit is worse than 20, 
the track that gives the largest contribution to the \Chisq\ is removed
and the vertex fit is repeated.
This procedure is iterated until the \Chisq\ requirement is satisfied.
Since the method does not properly treat 
displaced charm vertices and their daughter tracks,
a degradation of the vertex resolution and a bias on the vertex position
is introduced.   An
interaction point constraint is applied to the vertex fit 
for both $B$ mesons in order to improve the vertex resolution.
The typical $\delz$ resolution is 100 \micron.
The proper-time difference is approximated as $\dt\approx\delz/c(\beta\gamma)_\Upsilon$ where $(\beta\gamma)_\Upsilon$ is $\beta\gamma$ of the $\Upsilon(4S)$ in the laboratory frame.


The likelihood function for $\bdstlnu$ lifetime fit is defined as 
\begin{eqnarray*}
&&\hspace*{-0.7cm}L(\tbz, \tbm, \smis, \fmis, \sbg, \mubg, \lmbg, \flmbg)\\
&&=\prod_i\int_{-\infty}^{\infty}d(\dtp) [p_{SIG}^i(\dtp)+p_{BG}^i(\dtp)],\\
&&\hspace*{-0.7cm}p_{SIG}^i(\dtp)=(\fz^i\frac{e^{-\frac{|\dtp|}{\tbz}}}{2\cdot\tbz}+\fm^i\frac{e^{-\frac{|\dtp|}{\tbm}}}{2\cdot\tbm})\\\
&&[(1-\fmis)\frac{e^{-\frac{(\dtb-\dtp-\mu)^2}{2\sigdt^2}}}{\sqrt{2\pi}\sigdt}
+\fmis\frac{e^{-\frac{(\dtb-\dtp-\mumis)^2}{2(\sigmisdt)^2}}}{\sqrt{2\pi}\sigmisdt}],\\
&&\hspace*{-0.7cm}p_{BG}^i(\dtp)=\sum_k f_k^i\frac{e^{-\frac{(\dtb-\dtp-\mubg^k)^2}{2(\sbg^k\sigdt)^2}}}{\sqrt{2\pi}\sbg\sigdt}\\
&&[(1-\flmbg^k)\cdot\delta(\dtp)+\flmbg^k\frac{\lmbg^k}{2}e^{-\lmbg^k|\dtp|}],\\
\end{eqnarray*}
where: $\tbz$ and $\tbm$ are the \Bzb\ and \Bm\ lifetimes;
$\sigdt$ and $\sigmisdt$ are the main and tail parts 
of the $\dt$ resolution calculated event-by-event from the track 
error matrix as described below;
$\fmis$ denotes the fraction of the tail part of 
the signal resolution function and is determined from the fit.
$\mu$ and $\mumis$ are the biases due to the charm meson daughter tracks,
determined from the MC simulation;
$\sbg$, $\mubg^k$, $\lmbg^k$, $\flmbg^k$ are background-shape parameters, 
determined from the fit (fake \Dstar), 
data (fake lepton) or MC (random $\dstar\ell$);
$\fz^i$, $\fm^i$ and $f_k^i$ are fractions 
of the \Bzb\ and \Bm\ signals and background 
contributions that are calculated event-by-event using
the measured $\dM_{\dstar}$ value.
The $\bdstxlnu$ background fractions are estimated from
the known branching fractions and included in 
$\fz^i$ and $\fm^i$, since the effect of the 
missing $X$ is found to be negligible.

\begin{figure}[tbh]
\begin{center}
\resizebox{0.48\textwidth}{!}{\includegraphics{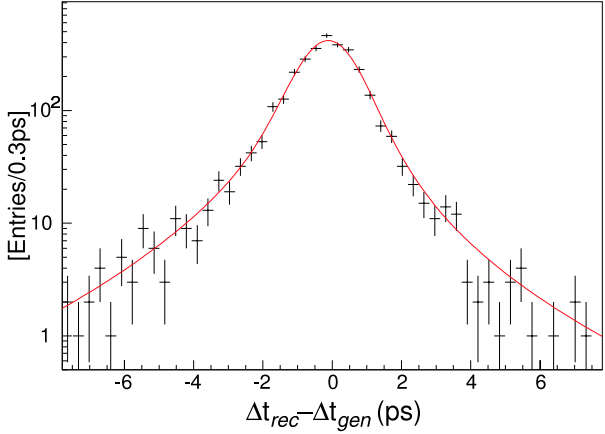}}
\caption{The $\dtrec-\dtgen$ distribution and 
resolution function for MC signal events.}
\label{fig:res}
\end{center}
\end{figure}

The $\dt$ resolution is a convolution of 
the $\delz$ resolution and the error 
due to the kinematic approximation 
($\dt\approx\delz/c(\beta\gamma)_\Upsilon$) $\sigk$:
$$ \sigdt^2=[\sigdz/c(\beta\gamma)_\Upsilon]^2+\sigk^2.$$
The $\delz$ resolution $\sigdz$ is 
calculated from the vertex resolutions of the 
reconstructed ($\sigzrec$) and associated ($\sigzasc$) $B$ mesons:
$$\sigdz^2=(\sdet\sigzrec)^2+(\sdet^2+\scharm^2)(\sigzasc)^2,$$
where $\sdet$ is a global 
scaling factor that accounts for any systematic bias in 
the resolution calculation from the track-helix errors, 
and $\scharm$ is a scaling factor to account for 
the degradation of the vertex resolution of the 
associated $B$ meson due to contamination of charm daughters.
If the reduced \Chisq\ (${\chi}^2/n$) of the vertex fit 
is worse than 3, the corresponding vertex error 
($\sigzrec$ or $\sigzasc$) is scaled by $[1+\alpha({\chi}^2/n-3)]$.
This ${\chi}^2/n$-dependent scaling is essential 
to account for events with large errors.
We use the value of $\sdet=0.99\pm0.04$ determined from
the \Dz\ lifetime fit in the $z$ direction.  The values for
$\sigk$, $\scharm$ and $\alpha$ are determined from the MC.
$\sigmisdt$ is calculated in a similar manner.
The associated parameter $\smis$ is determined in the fit along with $\fmis$.
Figure \ref{fig:res} shows the $\dtrec-\dtgen$ distribution and 
resolution function for MC signal events.

The likelihood function for the hadronic modes is defined as 
\begin{eqnarray*}
&&\hspace*{-0.7cm}L(\tsig, \sbg, \smisbg, \mubg, \mumisbg, \fmisbg, \lmbg, \flmbg)\\
&&=\prod_i\int_{-\infty}^{\infty}d(\dtp) [p_{SIG}^i(\dtp)+p_{BG}^i(\dtp)],\\
&&\hspace*{-0.7cm}p_{SIG}^i(\dtp)=(1-\fbg^i)\frac{e^{-\frac{|\dtp|}{\tsig}}}{2\cdot\tsig}\\\
&&[(1-\fmis)\frac{e^{-\frac{(\dtb-\dtp-\mu)^2}{2\sigdt^2}}}{\sqrt{2\pi}\sigdt}
+\fmis\frac{e^{-\frac{(\dtb-\dtp-\mumis)^2}{2(\sigmisdt)^2}}}{\sqrt{2\pi}\sigmisdt}],\\
&&\hspace*{-0.7cm}p_{BG}^i(\dtp)=\fbg^i[(1-\fmisbg)\frac{e^{-\frac{(\dtb-\dtp-\mubg)^2}{2(\sbg\sigdt)^2}}}{\sqrt{2\pi}\sbg\sigdt}\\
&&+\fmisbg\frac{e^{-\frac{(\dtb-\dtp-\mumisbg)^2}{2(\smisbg\sigdt)^2}}}{\sqrt{2\pi}\smisbg\sigdt}]\\
&&[(1-\flmbg)\cdot\delta(\dtp)+\flmbg\frac{\lmbg}{2}e^{-\lmbg|\dtp|}].\\
\end{eqnarray*}
The fraction of background  $\fbg^i$ is calculated from
the \DE\ and \Mb\ values for each event.  
The background shape parameters $\sbg$, $\smisbg$, $\mubg$, $\mumisbg$, $\fmisbg$, $\lmbg$ and $\flmbg$ are determined from the fit.
We use $\sdet=0.94\pm0.04$ in the $\bpsik$ analysis to account for slightly different kinematic properties from $\bdstlnu,\ D\pi$ decays.

Figure \ref{fig:fit} shows the $\dt$ distributions and 
fit results for $\bzdstlnu$ and $\bmpsikm$ events.
Table \ref{table:results} summarizes the measurement results. 
The main sources of  systematic errors are
uncertainties in the resolution function 
and the $\dt$ dependence of the reconstruction efficiency. 
All results are preliminary.
\begin{figure}[bh]
\begin{center}
\resizebox{0.48\textwidth}{!}{\includegraphics{lifetime_fit.epsi}}
\caption{The $\dt$ distributions and fit results 
for (a) $\bzdstlnu$ and (b) $\bmpsikm$ events.  The
dotted curve represents the background.}
\label{fig:fit}
\end{center}
\end{figure}

\newpage

\begin{table}[htdp]
\caption{Summary of lifetime measurements.}
(a) $B$ lifetime measurements.
\begin{center}
\begin{tabular}{|c|c|}
\hline
$\bzdstlnu$  & ($\tbzdstlnu$) ps\\
$\bzdstpm$   & ($\tbzdstpm$) ps \\
$\bzdppm$    & ($\tbzdppm$) ps  \\
$\bzbpsikst$ & ($\tbzpsikst$) ps\\ \hline
$\bzb$ combined     & ($\tbzresult$) ps\\ \hline
$\bzbpsiks$  & ($\tbzpsiks$) ps \\\hline\hline
$\bmdstlnu$  & ($\tbmdstlnu$) ps\\
$\bmdzpm$    & ($\tbmdzpm$) ps  \\
$\bmpsikm$   & ($\tbmpsikm$) ps \\ \hline
$\bm$ combined     & ($\tbmresult$) ps\\ \hline\hline
$\rbm$ & $\rbmresult$ \\
$\ycp$ & $\yresult$ \\ \hline
\end{tabular}
\end{center}
\vspace{1.3ex}
(b) $D$ lifetime measurements.
\begin{center}
\begin{tabular}{|c|c|}
\hline
$\dzkpi$ & ($\tdzresult$) fs\\
$\dzkk$ & ($410.5 \pm 14.3 ^{+9.7}_{-5.9}$) fs\\ \hline\hline
$\dpkpipi$ & ($1049 ^{+25}_{-24} {}^{+16}_{-19}$) fs \\
$\dpphipi$ & ($974^{+68}_{-62} {}^{+26}_{-18}$) fs\\ \hline
\Dp\ combined & ($\tdpresult$) fs \\ \hline\hline
$\dsphipi$ & ($470 \pm 19 ^{+5}_{-7}$) fs \\
$\dskstk$ & ($505 ^{+34}_{-33} {}^{+8}_{-12}$) fs \\ \hline
\Ds\ combined & ($\tdsresult$) fs \\ \hline\hline
$\rdp$ & $\rdpresult$ \\
$\rds$ & $\rdsresult$ \\ 
$\ycp$ & ($\ydresult$) \% \\ \hline
\end{tabular}
\end{center}
\label{table:results}
\end{table}%

\end{document}